\documentclass[prb,showpacs,twocolumn,preprintnumbers]{revtex4}

\usepackage[dvips]{graphics}
\usepackage{amsmath, amsthm, amssymb}
\usepackage{graphicx}
\usepackage{dcolumn}
\usepackage{bm}

\newcommand{\tr}{\mathrm{Tr}}

\newcommand{\eg}{{\it{e.g.}}, }
\newcommand{\ie}{{\it{i.e.}}, }
\newcommand{\etal}{{\it{et al.}}, }

\usepackage{dcolumn}
\usepackage{bm}

\begin{document}

\title{Entanglement quantifiers and phase transitions}

\author{D. Cavalcanti}\email{dcs@fisica.ufmg.br}
\affiliation{Departamento de F\'{\i}sica - Caixa Postal 702 -
Universidade Federal de Minas Gerais - 30123-970 - Belo Horizonte -
MG - Brazil}
\author{F.G.S.L. Brand\~ao}\email{fernando.brandao@imperial.ac.uk}
\affiliation{QOLS, Blackett Laboratory, Imperial College London,
London SW7 2BW, UK}\affiliation{Institute for Mathematical Sciences,
Imperial College London, London SW7 2BW, UK}
\author{M.O. \surname{Terra Cunha}}\email{tcunha@mat.ufmg.br}
\affiliation{Departamento de Matem\'atica - Caixa Postal 702 -
Universidade Federal de Minas Gerais - 30123-970 - Belo Horizonte -
MG - Brazil}

\begin{abstract}
By the topological argument that the identity matrix is surrounded
by a set of separable states follows the result that if a system is
entangled at thermal equilibrium for some temperature, then it
presents a phase transition (PT) where entanglement can be viewed as
the order parameter. However, analyzing several entanglement
measures in the 2-qubit context, we see that distinct entanglement
quantifiers can indicate different orders for the same PT. Examples
are given for different Hamiltonians. Moving to the multipartite
context we show necessary and sufficient conditions for a family of
entanglement monotones to attest quantum phase transitions.
\end{abstract}

\pacs{03.65.Ud,64.60.-i,73.43.Nq}

\maketitle

\section{Introduction}

The study of phase transitions under the view of exclusively quantum
correlations has hooked the interest of the quantum information
community recently \cite{ON,ABV}. Linking entanglement and (quantum)
phase transitions (PTs) is tempting since PTs are related to
correlations of long range among the system's constituents
\cite{Yang}. Thus expecting that entanglement presents a peculiar
behavior near criticality is natural.

Recent results have shown a narrow connection between entanglement
and critical phenomena. For instance, bipartite entanglement has
been widely investigated near to singular points for exhibit
interesting patterns \cite{ON,ABV}. The {\emph{localizable
entanglement}} \cite{LocEnt} has been used to show certain critical
points that are not detected by classical correlation functions
\cite{VDC04}. The {\emph{negativity}} and the {\emph{concurrence}}
quantifiers were shown to be quantum-phase-transitions witnesses
\cite{LA}. Furthermore, closely relations exist between entanglement
and the order parameters associated to the transitions between a
normal conductor and a superconductor and between a Mott-insulator
and a superfluid \cite{Brab}.

The main route that has been taken in order to capture these ideas
is through the study of entanglement in specific systems. However it
is believed that a more general picture can be found. Here, we go
further in this direction starting from the generic result that, for
a bipartite system at thermal equilibrium with a reservoir, there
exist two distinct phases, one in which some entanglement is present
and another one where quantum correlations completely vanish. We
then exemplify this result with 2-qubit systems subjected to
different Hamiltonians and curiously it is viewed that, by choosing
different entanglement quantifiers , one attributes different orders
to the phase transition.

Although multipartite entanglement also plays an important role in
many-body phenomena (its is behind some interesting effects such as
the \emph{Meissner effect} \cite{Meis}, the {\emph{high-temperature
superconductivity}} \cite{high-Tc}, and \emph{superadiance}
\cite{Dicke}), rare results linking it to PTs exist. Crossing this
barrier is also a goal of this Letter. For that, we give necessary
and sufficient conditions to a large class of multipartite
entanglement quantifiers to signal singularities in the ground state
energy of the system. We finish this work discussing a recently
introduced quantum phase transition, the \emph{geometric phase
transition}, which takes place when  a singularity in the boundary
of the set of entangled states exists.

A phase transition occurs when some state function of a system
presents two distinct phases, one with a non-null value and another
one in which this function takes the null value \cite{LL}. Such a
function is called an {\emph{order parameter}} for the system.
However one can think that this is a very tight definition and want
to define a PT as a singularity in some state function of the system
due to changes in some parameter (coupling factors in the
Hamiltonian, temperature, etc). By extension, this function is also
called the order parameter of the PT\footnote{Sometimes the order
parameter is not a measurable property of the system, but we do not
want to enter into this merit.}. Note that the first definition of
PT is a special case of the latest one. When the singularity
expresses itself as a discontinuity in the order parameter we say
that we are dealing with a discontinuous PT. If the discontinuity
happens in some of the derivatives of the order parameter, say the
$n^{th}$-derivative, it is said to be a $n^{th}$-order PT, or a
continuous PT. In this paper we will consider entanglement as a
state function and see that it can present a singularity when some
parameter of the problem changes. Thus, we make a more general
discussion about when a given entanglement quantifier, or some of
its derivatives, can present a discontinuity.

\section{The entangled$\rightarrow$disentangled transition}

The first phase transition we will discuss is when a system is in
thermal equilibrium with a reservoir. This system can show two
phases: one separable and other entangled. The following question
raises: is this transition smooth? We will show that the answer for
this question depends on the entanglement quantifier adopted.

Let us first revisit a very general result following just from a
topological argument. Given a quantum system with Hamiltonian $H$,
its thermal equilibrium state is given by $\rho(T)=\frac{\exp(-\beta
H)}{Z}$, where $Z=\tr \exp(-\beta H)$ is the partition function and
$\beta=(k_B T)^{-1}$, $k_B$ denoting the Boltzmann constant and $T$
the absolute temperature. This state is a continuous function of its
parameters. If the space state of the system has finite dimension
$d$, then $\lim_{T\rightarrow\infty}\rho(T)=\frac{I}{d}$, where $I$
denotes the identity operator. For multipartite systems,
$\frac{I}{d}$ is an interior point in the set of separable states
\cite{ZHSL}, \ie it is separable and any small perturbation of it is
still separable. The thermal equilibrium states $\rho(T)$ can be
viewed as a continuous path on the density matrix operators set,
ending at $\frac{I}{d}$. So if for some temperature $T_e$ the state
$\rho(T_e)$ is non-separable, there is a finite critical temperature
$T_c>T_e$ such that $\rho(T_c)$ is in the boundary of the set of
separable states. An important class of examples is given by the
systems with entangled ground state\footnote{Bipartite systems with
factorizable ground states can have thermal equilibrium states
separable for all temperatures, or can also show entanglement at
some temperature. In this case, there will be (at least) two phase
transitions when temperature is raised: one from separable to
entangled, and another from entangled to separable \cite{ABV}. Also
multipartite versions of this theorem can be stated: for each kind
of entanglement which the system shows at some temperature, there
will be a finite temperature of breakdown of this kind of
entanglement.}, \ie $T_e=0$.

It is clear that the entanglement $E$ of the system will present a
singularity at $T_c$. Thus $E$ can be viewed as a true order
parameter in the commented PT. Moreover let us explore a little bit
more the result that ``thermal-equilibrium entanglement vanishes at
finite temperature''\cite{FMB,Rag}. It will be shown that different
entanglement quantifiers attribute different orders for this PT. For
that we will show an entanglement quantifier that is discontinuous
at $T_c$, two others presenting a discontinuity at its first
derivative (asserting a $1^{th}$-order PT), and another one in which
the discontinuity manifests itself in
$\frac{d^{2}E(\rho)}{dT^2}|_{T=T_c}$ (asserting a $2^{nd}$-order
PT).

As the first example take the Indicator Measure, $IM(\rho)$, defined
as 1 for entangled states and 0 for separable ones. Although $IM$ is
an entanglement monotone\footnote{Entanglement monotones are
quantifiers that do not increase when LOCC-operations are applied in
$\rho$ \cite{GV}. This feature has been viewed by many people as the
unique requirement for a good quantifier of entanglement.} it is
quite weird once it is a discontinuous function itself. Of course
$IM$ presents a discontinuity at $T=T_c$, \ie when $\rho$ crosses
the border between the entangled and the disentangled-states world.

However it is interesting to study some best behavioured and
well-accepted entanglement monotones, and we will do that through
some examples in the 2-qubit context. Take the concurrence $C$, the
entanglement of formation $E_f$ and the negativity $N$. These three
functions are able to quantify entanglement properly although, as it
will be seen, in different manners. The entanglement of formation
was proposed by Bennett {\it{et al.}}\cite{Ef} as the infimum of
mean pure state entanglement among all possible ensemble
descriptions of a mixed state $\rho$. The concurrence was developed
by Wootters and collaborators \cite{Woo} in the context of trying to
figure out a feasible way to calculate the entanglement of
formation. Thus $E_f$ and $C$ are connected by
\begin{equation}\label{Ef(C)}
E_f(\rho) =
H_{2}\left(\frac{1}{2}+\frac{1}{2}\sqrt{1-C^2(\rho)}\right),
\end{equation}
where $H_{2}(x)=-x\log x -(1-x)\log(1-x)$ and it is assumed that
$0\log 0=0$. The concurrence can be defined by
\begin{equation}
C(\rho)=\max(0,\lambda_1-\lambda_2-\lambda_3-\lambda_4),
\end{equation}
with $\lambda_i$ being the square roots of the eigenvalues of the
matrix
$\rho(\sigma_{y}\otimes\sigma_{y})\rho^{*}(\sigma_{y}\otimes\sigma_{y})$
in decreasing order and $\sigma_{y}$ is the Pauli matrix.

On the other hand the negativity uses the idea of partial transpose
to calculate entanglement \cite{VW,ZHSL}. It can be defined as
\begin{equation}
N(\rho)=\|\rho^{T_{A}}\|-1,
\end{equation}
where the subscript $T_A$ indicates the partial transpose operation
and $\|\star\|$ means the trace norm. Alternatively, one can define
the logarithmic negativity as \cite{VW,ZHSL}
$E_{\mathcal{N}}(\rho)=\log_{2}(1+N(\rho))$.

\begin{figure}\centering
   \rotatebox{270}{\includegraphics[scale=0.30]{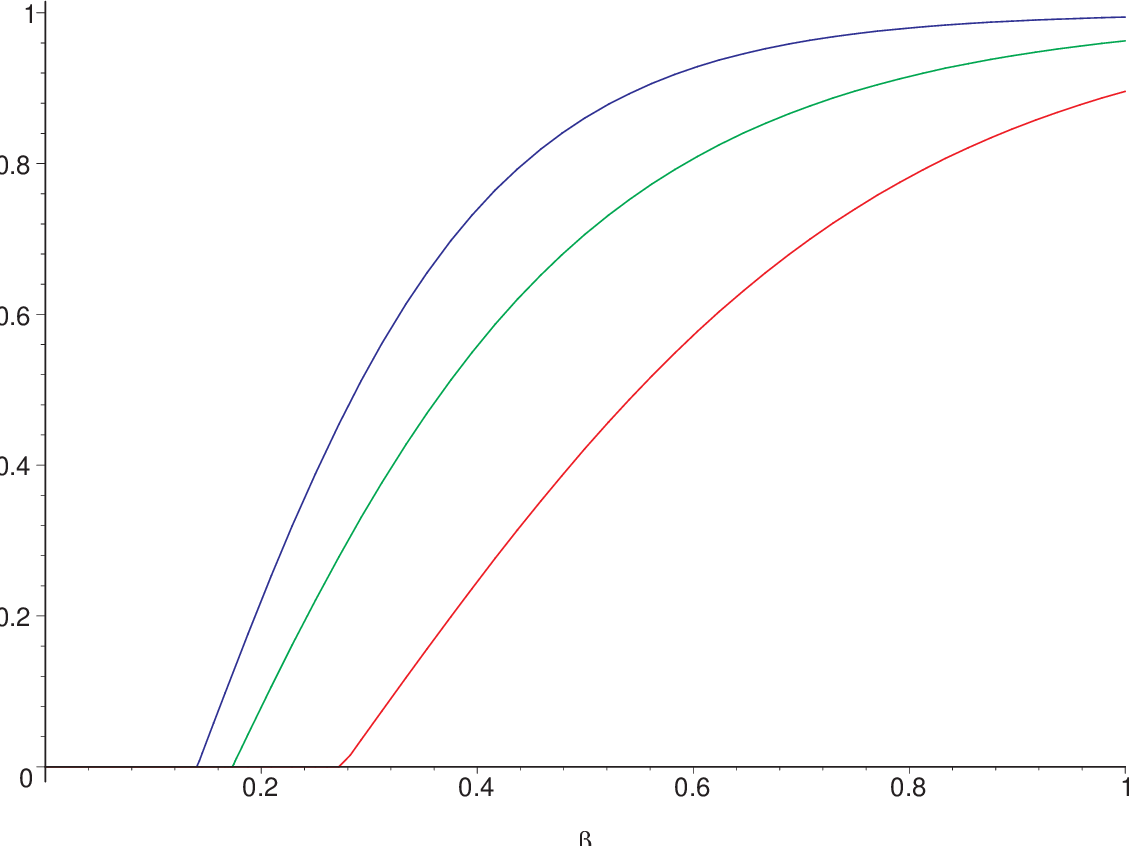}} \\
    \rotatebox{270}{\includegraphics[scale=0.30]{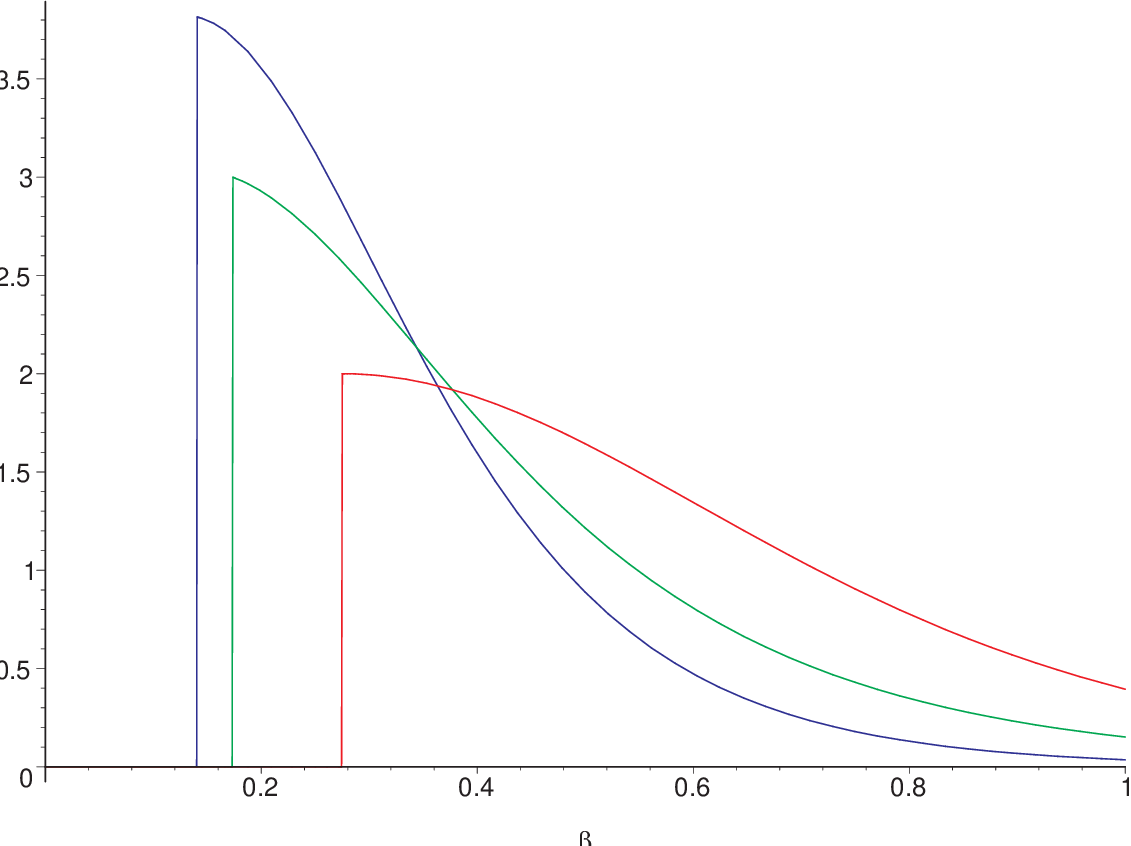}}
\caption{(Color online) {\textbf{Above:}} $C(\rho)$ vs. $\beta$ for
$x=1,y=1,z=1$ (red); $x=3,y=1,z=1$
  (green); and
  $x=3,y=2,z=1$ (blue). {\textbf{Below:}}$\frac{dC(\rho)}{d\beta}$ vs.
  $\beta$ for the same values of $x, y,$ and $z$.
   $C$ shows a PT of $1^{st}$ order (its 1st derivative is discontinuous). In the cases considered
   $C(\rho)$=$N(\rho)$,
   and the conclusions are also valid for the negativity\cite{Ver}.}
\label{C}\end{figure}

\begin{figure}\centering
\rotatebox{270}{\includegraphics[scale=0.30]{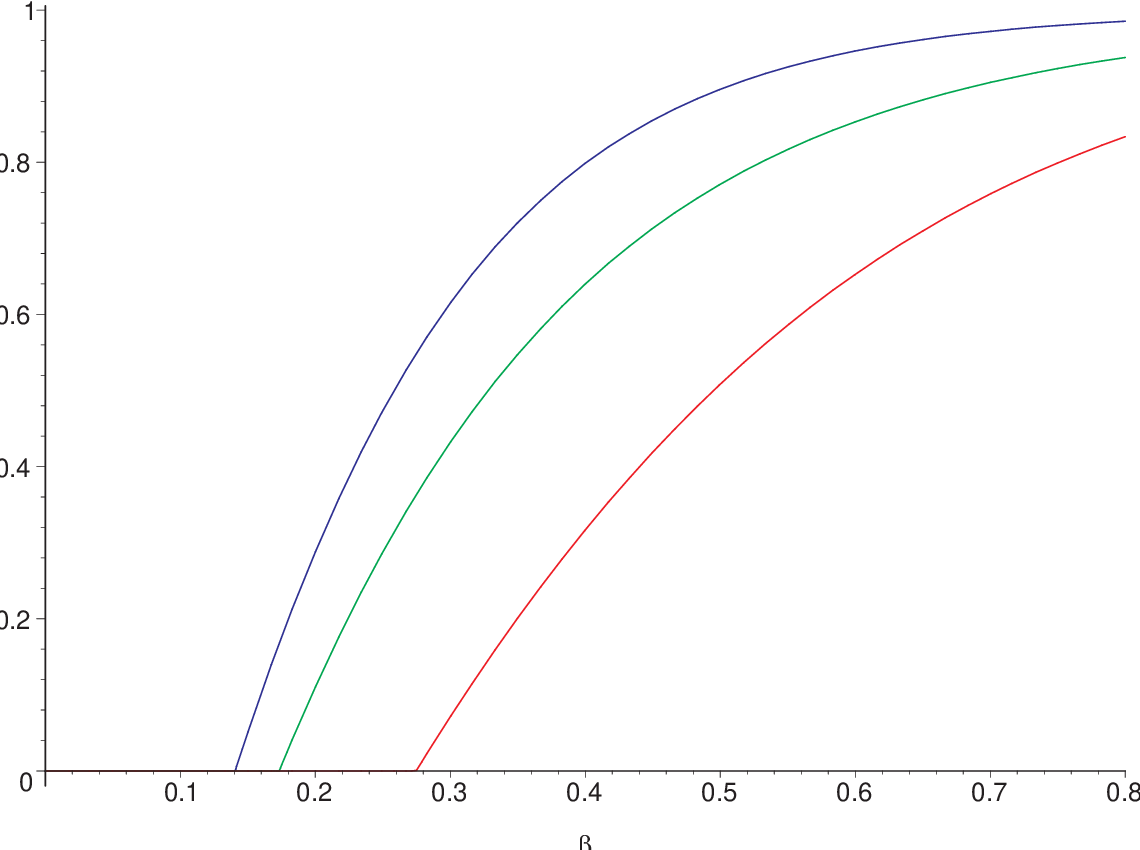}} \\
\rotatebox{270}{\includegraphics[scale=0.30]{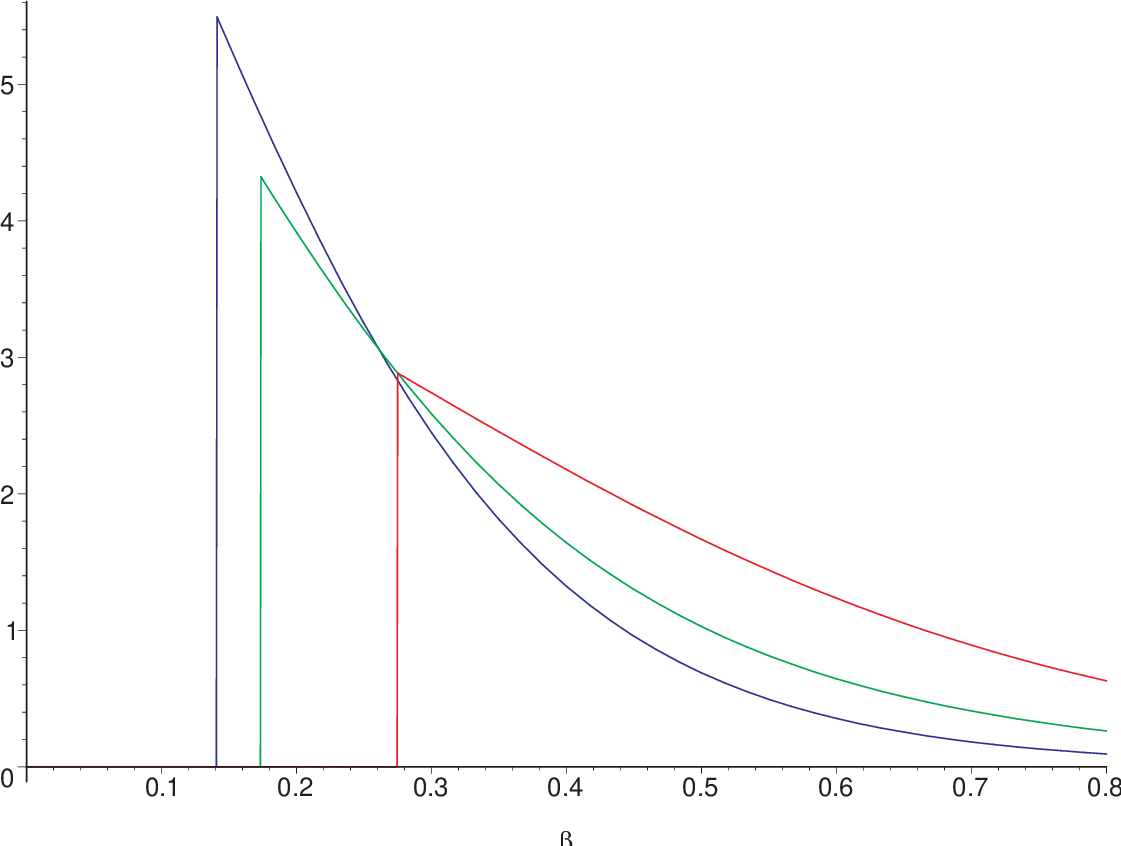}} \\
   \caption{(Color online) {\textbf{Above:}} $E_{\mathcal{N}}(\rho)$ vs. $\beta$ for $x=1,y=1,z=1$ (red); $x=3,y=1,z=1$ (green);
  and $x=3,y=2,z=1$ (blue). {\textbf{Below:}} $\frac{dE_{\mathcal{N}}(\rho)}{d\beta}$ vs.
  $\beta$ for the same values of $x, y,$ and $z$.}
\label{LN}\end{figure}

\begin{figure}\centering
 \rotatebox{270}{\includegraphics[scale=0.30]{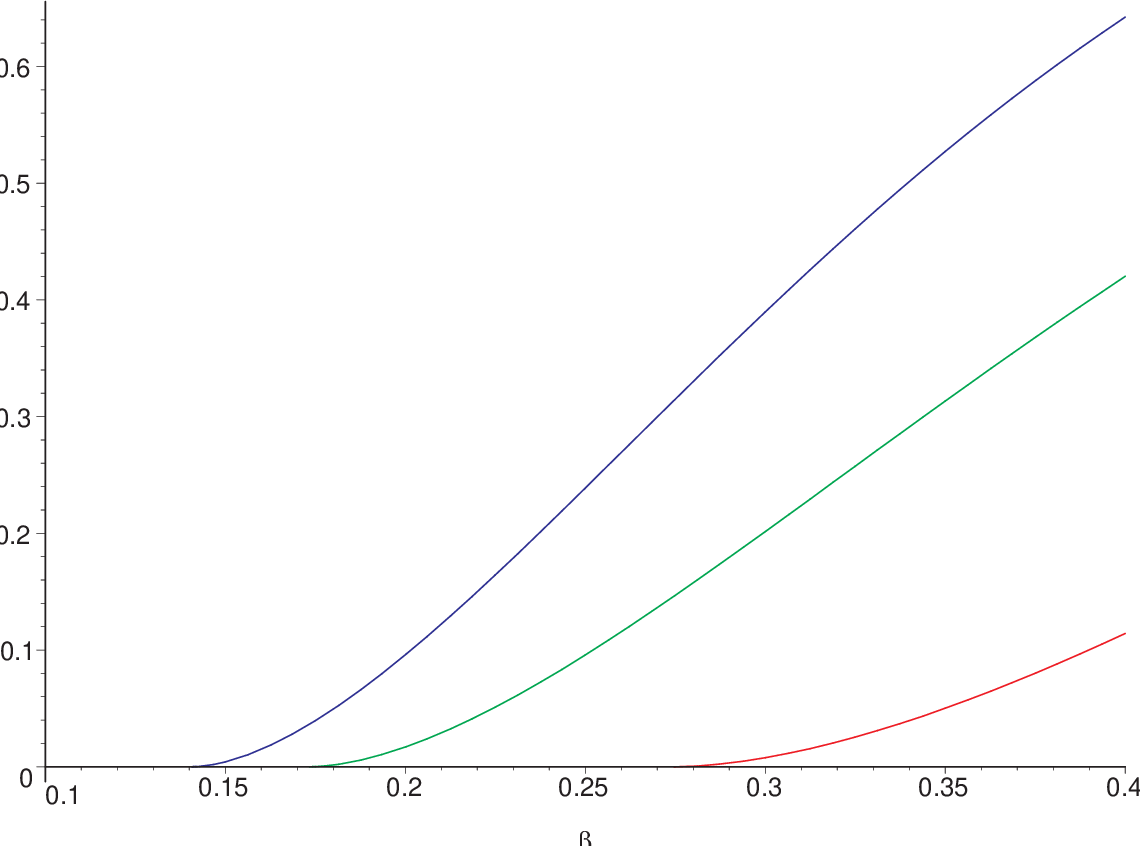}}\\
 \rotatebox{270}{\includegraphics[scale=0.30]{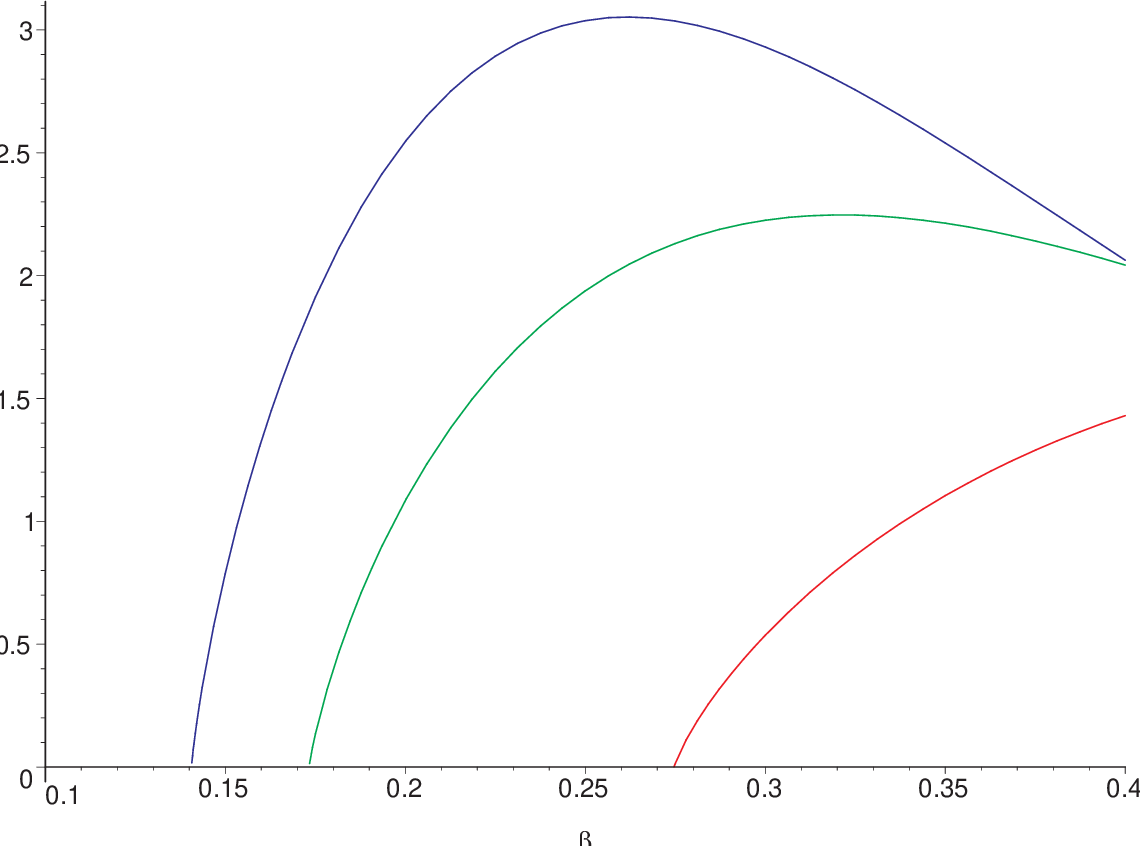}}
\caption{(Color online) {\textbf{Above:}} $E_f(\rho)$ vs. $\beta$
for $x=1,y=1,z=1$ (red); $x=3,y=1,z=1$ (green); and $x=3,y=2,z=1$
(blue). {\textbf{Below:}}$\frac{dE_f(\rho)}{d\beta}$ vs. $\beta$ for
the same values of $x, y,$ and $z$.} \label{Ef}\end{figure}

Let us use these quantifiers to study the entanglement of
thermal-equilibrium states,
\begin{equation}\rho=\frac{\exp(-\beta H)}{Z},
\label{rho}\end{equation} subject to a completely non-local
Hamiltonian of the form \cite{Dur+}
\begin{equation}H=x\sigma_x\otimes\sigma_x+y\sigma_y\otimes\sigma_y+z\sigma_z\otimes\sigma_z.
\label{H}\end{equation} Note that the 1D 2-qubit Heisenberg chain is
a particular case of \eqref{H} when $x=y=z=J$ ($J<0$ being the
ferromagnetic and $J>0$ the antiferromagnetic cases). The results
are plotted in Figs. \ref{C}, \ref{LN}, and \ref{Ef}.

An interesting conclusion following from the figures is that
according to $E_F$ the transition is of $2^{nd}$ order, according to
$C$ (and $N$ as well) and $E_{\mathcal{N}}$ it is of $1^{st}$ order,
and remember that, according to $IM$ all transitions are
discontinuous. In fact it is possible to see, directly from its
definition, that $E_{\mathcal{N}}$ will always present a
discontinuity in the same derivative as $N$. For this aim we can
write:
\begin{equation}
\frac{dE_{\mathcal{N}}(\beta)}{d\beta}=\frac{1}{(1+N(\beta))\ln2}\frac{dN(\beta)}{d\beta}.
\end{equation}
Similarly, the relation between $E_f$ and $C$ can be also verified
analytically. The derivative of $E_f$ with respect to $\beta$ is
\begin{equation}
\frac{dE_f(\beta)}{d\beta}=\frac{C(\beta)}{2\sqrt{1-C^2(\beta)}}
\log\left(\frac{1-\sqrt{1-C^{2}(\beta)}}{1+\sqrt{1-C^{2}(\beta)}}\right)\frac{dC(\beta)}{d\beta}.
\end{equation}
So it is possible to see that, even $C(\beta)$ being singular at
$\rho_c$ (it is, when $T=T_c$), the singularity manifests itself on
$E_f(\beta)$ only to the next order.

In fact, this situation resembles that in percolation theory, when different
``percolation quantifiers'' like probability of percolation, the mean size of the clusters,
and the conductivity between two points show different critical behaviour\cite{SA}.

\section{Multipartite entanglement as indicator of quantum phase transitions}

In Ref.\cite{LA}, the authors show that the concurrence and the
negativity serve themselves as quantum-phase-transition indicators.
This is because, unless artificial occurrences of non-analyticities,
these quantifiers will present singularities if a quantum phase
transition happens. An extend result for another bipartite
entanglement quantifiers is presented in ref.\cite{LA2}. In the same
context, Rajagopal and Rendell offer generalizations of this theme
to the more general case of mixed state\cite{RR05}.

By following the same line of research we now extend the previous
results to the multipartite case. We will see that it is possible to
establish some general results, similar to Ref.\cite{LA}, also in
the multipartite scenario. We can use for this aim the
\emph{Witnessed Entanglement}, $E_W(\rho)$, to quantify entanglement
\cite{FB} (this way of quantifying entanglement includes several
entanglement monotones as special cases, such as the robustness and
the best separable approximation measure). Before giving the
definition of $E_W$ we must review the concept of entanglement
witnesses. For all entangled state $\rho$ there is an operator that
witnesses its entanglement through the expression $\tr(W\rho)<0$
with $\tr(W\sigma)\geq0$ for all $k$-separable states $\sigma$ (we
call $k$-separable every state that does not contain entanglement
among any $m>k$ parts of it, and denote this set $S_k$) \cite{FB}.
We are now able to define $E_W$. The witnessed entanglement of a
state $\rho$ is given by
\begin{equation}\label{Ew}
E_{W}^{k}(\rho)=\max \{0, -\min_{W\in\mathcal{M}}Tr(W\rho)\},
\end{equation}
where the choice of $\mathcal{M}$ allows the quantification of the
desired type of entanglement that $\rho$ can exhibit. The
minimization of $\tr(W\rho)$ represents the search for the optimal
entanglement witness $W_{opt}$ subject to the constraint
$W\in\mathcal{M}$. The interesting point is that by choosing
different $\mathcal{M}$, $E_W$ can reveal different aspects of the
entanglement geometry and thus quantify entanglement under several
points of view. As a matter of fact, if in the minimization
procedure in \eqref{Ew} it is chosen to search among witnesses $W$
such that $\tr(W)\leq I$ ($I$ is the identity matrix), $E_W$ is
nothing more than the {\emph{generalized robustness}}, an
entanglement quantifier \cite{VT} with a rich geometrical
interpretation \cite{Nos2,Nos3}. Other choices of $\mathcal{M}$
would reach other known entanglement quantifiers \cite{FB}. Moreover
it is easy to see that, regardless these choices, $E_W$ is a
bilinear function of the matrix elements of $\rho$ and of $W_{opt}$.
So singularities in $\rho$ or in $W_{opt}$ cause singularities in
$E_W(\rho)$.

At this moment we can follow Wu \etal in Ref.~\cite{LA} and state
that, if some singularity occurring in $E_W$ is not caused by some
artificial occurrences of non-analyticity (\eg maximizations or some
other mathematical manipulations in the expression for $E_W$ - see
conditions a-c in Theorem 1 of Ref.~\cite{LA}), then a singularity
in $E_W$ is both necessary and sufficient to signal a PT. It is
important to note that the concept of PT considered by the authors
is not thermal equilibrium PT: the PT's discussed by them are that
linked with non-analyticities in the derivatives of the ground state
energy with respect to some parameters as a coupling constant. On
the other hand it is also important to highlight that our result
implies a multipartite version of theirs. Moreover, the use of $E_W$
to studying quantum phase transitions can result in a possible
connection between critical phenomena and quantum information, as
$E_W$ (via the robustness of entanglement) is linked to the
usefulness of a state to teleportation processes \cite{Brac}.

We can go further in the concept of a PT and study the cases where
$E_W$ presents a singularity. An interesting case is when a
discontinuity happens in $W_{opt}$ and not in $\rho$. This can
happen for example if the set $S_k$ presents a sharp shape,
situation in which occurs the recently introduced \emph{geometric
phase transition}\cite{Nos2}, where the PT is due the geometry of
$S_k$. Besides the interesting fact that a new kind of quantum phase
transition can occur, the geometric PT could be used to study the
entanglement geometry. This can be made by smoothly changing some
density matrix and establishing whether $E_W$ reveals some
singularity. Furthermore, $E_W$ can be experimentally evaluated, as
witness operators are linked with measurement processes
\cite{GHB,TG} and has been used to attest entanglement
experimentally \cite{BEK}. So, the geometry behind entanglement can
even be tested experimentally. A more detailed study of this issue
is given in Ref.\cite{Nos2}.
\begin{figure}
   \includegraphics[scale=0.40]{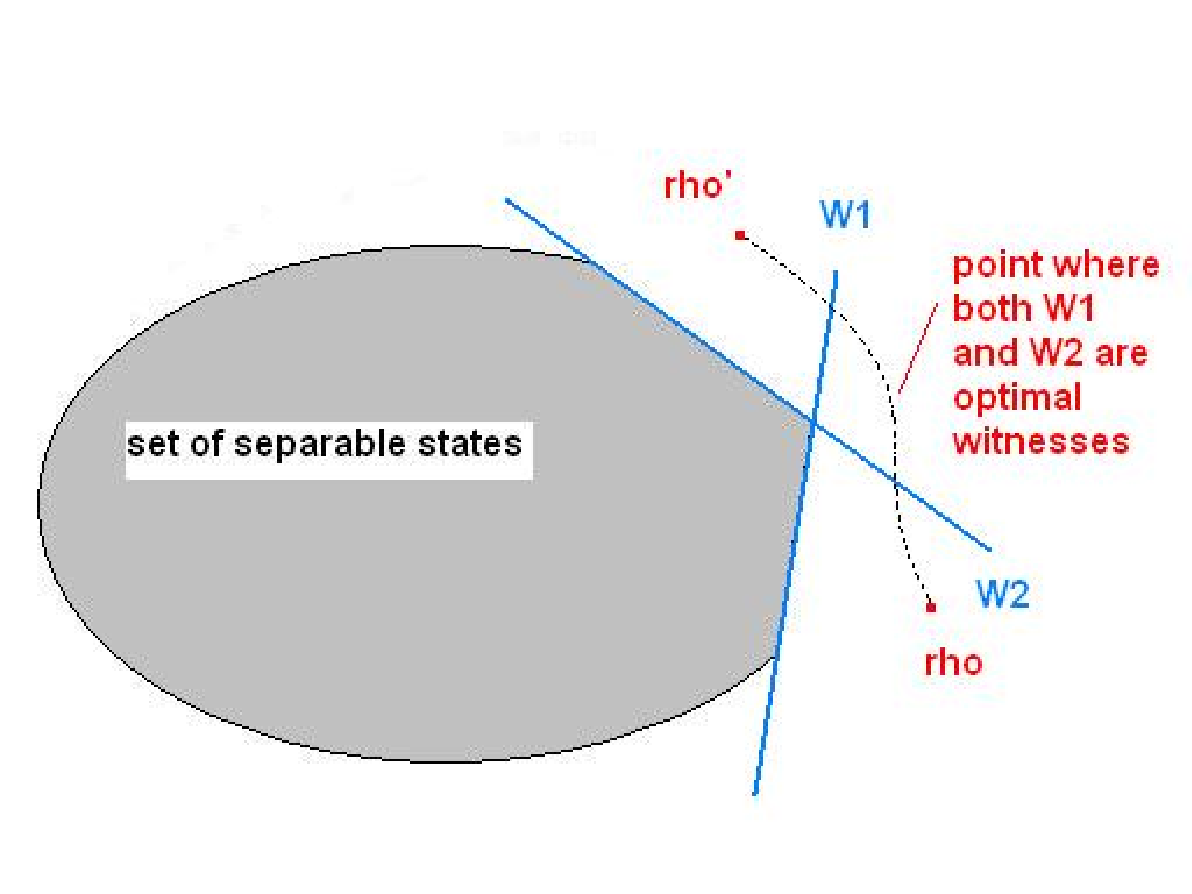}\\
  \caption{(Color online)The red (dot) line represents the way followed by $\rho$ when some
  parameter of the systems is changed. Geometrically, entanglement
  witness can be interpreted as tangent hyperplanes  to $S_k$. At a certain point both witnesses $W1$ and $W2$ are optimal for $\rho$.
  At this point there is a singularity in $E_W(\rho)=-Tr(W_{1 or 2}\rho)$.}\label{figEw}
\end{figure}

\section{Conclusion}

Summarizing, we have shown that entangled thermal-equilibrium
systems naturally present a phase transition when heated: the
entanglement-disentanglement transition. However different
entanglement quantifiers lead with this PT differently, in the sense
that, according to some of them the PT is of $1^{st}$-order (\eg the
negativity and concurrence), $2^{nd}$-order (\eg the entanglement of
formation), and even though discontinuous (\eg the indicator
measure). With these ideas in mind it is tempting to make some
questions: Is the PT showed here linked with some other physical
effect other than just vanishing quantum correlations? In other
words, which macroscopically observed PT have entanglement as order
parameter? Can the way in which entanglement quantifiers lead with
PT be considered a criterion for choosing among them? Is there ``the
good'' quantifier to deal with such PT? We hope our present
contribution can help in answering these questions.

Recent discussions have shown that the entanglement-disentanglement
transition is behind important quantum phase transitions
\cite{Brab}. So similar analysis can also be performed in different
contexts other than temperature increasing. Decoherence processes
could be a rich example.

\begin{acknowledgements}
We thank R. Dickman and V. Vedral for valuable discussions on the
theme. D.C. and F.G.S.L.B. thank financial support from CNPq-Brazil.
\end{acknowledgements}


\begin{thebibliography}{0}

\bibitem{ON}T. J. Osborne and M. A. Nielsen, \pra {\bf{66}}, 032110
(2002); A. Osterloh, L. Amico, G. Falci, and R. Fazio, Nature
{\bf{416}}, 608 (2002); J. I. Latorre, E. Rico, and G. Vidal, Quant.
Inf. and Comp. {\bf{4}}, 048 (2004); G. Vidal, J. I. Latorre, E.
Rico, and A. Kitaev, Phys. Rev. Lett. {\bf{90}}, 227902 (2003); T.
Roscilde, P. Verrucchi, A. Fubini, S. Haas, and V. Tognetti, \prl
{\bf{93}}, 167203 (2004); T. Roscilde, P. Verrucchi, A. Fubini, S.
Haas, and V. Tognetti, Phys. Rev. Lett. {\bf{94}}, 147208 (2005); D.
Larsson and H. Johannesson, Phys. Rev. Lett. {\bf{95}}, 196406
(2005).

\bibitem{ABV}M. C. Arnesen, S. Bose, and V. Vedral, Phys. Rev. Lett. {\bf{87}}, 017901
(2001).

\bibitem{Yang}C. N. Yang, Rev. Mod. Phys. {\bf{34}}, 694 (1962).

\bibitem{LocEnt}F. Verstraete, M. Popp , and J. I.
Cirac, \prl {\bf{92}}, 027901 (2004); M. Popp, F. Verstraete, M. A.
Mart\'in-Delgado, and J. I. Cirac, \pra {\bf{71}}, 042306 (2005).

\bibitem{VDC04}F. Verstraete, M. A. Mart\'in-Delgado, and J. I.
Cirac, \prl {\bf{92}}, 087201 (2004).

\bibitem{LA}L.-A. Wu, M. S. Sarandy, and D. A. Lidar, Phys. Rev. Lett. 93, 250404 (2004).

\bibitem{Brab} F. G. S. L. Brand\~ao, New J. Phys. {\bf{7}}, 254
(2005).
.

\bibitem{Meis}V. Vedral, e-print quant-ph/0410021.

\bibitem{high-Tc}V. Vedral, New J. Phys. {\bf{6}}, 102 (2004).

\bibitem{Dicke}N. Lambert, C. Emary, and T. Brandes, Phys. Rev. Lett. {\bf{92}}, 073602
(2004).

\bibitem{LL}L. D. Landau and E. M. Lifshitz, Statistical Physics (Addison-Wesley, Reading, Mass., 1969).

\bibitem{ZHSL}K. \.Zyczkowski, P. Horodecki, A. Sanpera, and M.
Lewenstein, Phys. Rev. A {\bf{58}}, 883 (1998).

\bibitem{FMB} B. V. Fine, F. Mintert, and A. Buchleitner, Phys. Rev. B {\bf{71}}, 153105
(2005).

\bibitem{Rag}G. A. Raggio, J. Phys. A: Math. Gen. {\bf{39}}, 617
(2006); O. Osenda and G. A. Raggio, \pra {\bf{72}}, 064102 (2005).

\bibitem{GV}G. Vidal,\jmo{} {47}, {355} {(2000)}.

\bibitem{Ef}C. H. Bennett, H. J. Bernstein, S. Popescu, and B. Schumacher Phys. Rev. A {\bf{53}}, 2046 (1996).

\bibitem{Woo}W. K. Wootters, Phys. Rev. Lett. {\bf{80}}, 2245 (1998).
S. Hill and W. K. Wootters, Phys. Rev. Lett. {\bf{78}}, 5022 (1997).

\bibitem{VW}J. Eisert, PhD thesis University of Potsdam, January 2001; G. Vidal and R.F.
Werner, Phys. Rev. A \textbf{65}, 032314 (2002); K. Audenaert, M. B.
Plenio and J. Eisert, Phys. Rev. Lett. \textbf{90}, 027901 (2003);
M. B. Plenio, Phys. Rev. Lett. 95, 090503 (2005).


\bibitem{Dur+}W. D\"{u}r, G. Vidal, J.I. Cirac, N. Linden, and S. Popescu,
 Phys. Rev. Lett. {\bf{87}}, 137901 (2001). C. H. Bennett \etal Phys. Rev. A {\bf{66}}, 012305 (2002).

\bibitem{Ver}F. Verstraete et. al., J. Phys. A: Math. Gen. {\bf{34}}, 10327 (2001).

\bibitem{SA}D. Stauffer and A. Aharony, {\it{Introduction to Percolation Theory}}
(Taylor and Francis, 1992).

\bibitem{LA2}L.-A. Wu, M. S. Sarandy, D. A. Lidar, and L. J. Sham, eprint quant-ph/0512031.

\bibitem{RR05}A. K. Rajagopal and R. W. Rendell, eprint
quan-ph/0512102.

\bibitem{FB} F. G. S. L. Brand\~ao, Phys. Rev. A {\bf{72}}, 022310 (2005).

\bibitem{VT}M. Steiner. Phys. Rev. A \textbf{67}, 054305 (2003). G.
Vidal and R. Tarrach. Phys. Rev. A \textbf{59}, 141 (1999).

\bibitem{Nos2} D. Cavalcanti, F.G.S.L. Brand\~ao, and M.O. Terra Cunha, e-print quant-ph/0510068.

\bibitem{Nos3}D. Cavalcanti, F.G.S.L. Brand\~ao, and M.O. Terra
Cunha, \pra {\bf 72}, 040303(R) (2005).

\bibitem{Brac}F. G. S. L. Brand\~ao, e-print quant-ph/0510078. Ll.
Masanes, e-print quant-ph/0508071.

\bibitem{GHB}O. G\"{u}hne \etal Phys. Rev. A
{\textbf{66}}, 062305 (2002).

\bibitem{TG} G. T\'oth and O. G\"uhne, Phys. Rev. Lett. {\textbf{94}}, 060501 (2005).

\bibitem{BEK}M. Bourennane \etal \prl \textbf{92}, 087902 (2004); M. Barbieri \etal \prl
{\bf{91}} 227901 (2003).


\end{thebibliography}
\end{document}